# On the Application of Agile Project Management Techniques, V-Model and Recent Software Tools in Postgraduate Theses Supervision

Pouria Sarhadi*, Wasif Naeem**, Karen Fraser***, David Wilson****

*School of Physics, Engineering and Computer Science, University of Hertfordshire, Hatfield, UK, (e-mail: p.sarhadi@herts.ac.uk)
**School of Electronics, Electrical Engineering and Computer Science, Queen's University Belfast, Belfast, UK, (e-mail: w.naeem@ee.qub.ac.uk)
*** Centre for Educational Development, Queen's University Belfast, Belfast, UK, (e-mail: k.fraser@qub.ac.uk)
**** The Institute of Electronics, Communications and Information Technology (ECIT), Queen's University Belfast, Belfast, UK, (e-mail: d.wilson@qub.ac.uk)

**Abstract**: Due to the nature of most postgraduate theses in control engineering and their similarities to industrial and software engineering projects, invoking novel project control techniques could be effective. In recent decades, agile techniques have attracted popularity thanks to their attributes in delivering successful projects. Hence exploiting those methods in education and thesis supervision of engineering topics can facilitate the process. On the other hand, because of the limitations imposed by the CoVid19 pandemic, the integration of well-established online tools in collaborative education is noteworthy. This paper proposes an application of the agile project management method for the supervision of postgraduate students' theses in the general field of engineering. The study extends a Scrum technique combined with approved systems engineering and team working tools such as Jira Software, Microsoft Teams, and Git version control (Github website). A custom designed V-model to nail an outstanding thesis is presented. The overall blended method is beneficial to provide feedback and self-assessment aid for the students and the supervisors. Employing this technique has shown promising progress in easing the supervision of students whilst helping them to manage their projects.

*Keywords*: Control education, postgraduate thesis supervision, Agile Scrum project management, Jira software, online supervision, control engineering, systems engineering, blended supervision.

## 1. INTRODUCTION

Postgraduate thesis supervision is a long-term iterative process that needs appropriate concentration and mental effort. In particular, when the number of students increases or the supervisor involves with other administrative, research and teaching responsibilities, the supervision quality may suffer Based on three separate post-2000 surveys conducted in Sweden and Norway, about 23, 29, and 33% of students reported poor advisory, which in some cases led to students withdrawing from their studies (Tengberg, 2015). The supervisor should provide continuous feedback as part of an effective practice; however, increased student numbers have become a common challenge (Brodtkorb, 2019). Utilising a systematic formal process in advising the students, which includes holding regular meetings, participation of supervisor, defining the required tasks, and providing feedback during the lifecycle of a thesis is a possible solution. This could also contribute to recent efforts to improve the quality assurance of control engineering education (Goodwin, et al., 2010, Rossiter, 2020, Rossiter et al., 2020, Rossiter et al., 2021). Hence, novel approaches in thesis supervision are sought. Owing to the affinity between control engineering and software/systems engineering, state-of-the-art approaches of the latter can be adapted to control education. Accordingly, this paper proposes a combination of agile project management techniques, systems engineering (V-model), and recent software tools in postgraduate theses supervision.

A number of studies have investigated challenges and solutions to facilitate the students' learning during their theses research (Vilkinas, 2008, de Kleijn et al., 2012, Ghadirian, et al, 2014, Tengberg, 2015, Karunaratne, 2018, Brodtkorb, 2019). In Ghadirian, et al. (2014), "supervisory knowledge and skills", "atmosphere", "bylaws and regulations relating to supervision" and "monitoring and evaluation" are suggested as the main difficulties. Karunaratne (2018) has proposed blended Information Communications Technology (ICT) based approaches to improve the quality and effectiveness of the thesis projects. "scheduling/conducting meetings and activities", "delivery/exchange of relevant information", "punctuality in providing feedback and other necessary information", "transparency in communication" and students' self-assessment are identified as the main factors to deliver a successful thesis. De Kleijn et al., (2012) highlights the importance of personal support and supervisors' involvement in their students' theses. Despite the considerable efforts to enhance the quality of education in control engineering (Goodwin, et al., 2010, Rossiter, 2020, Rossiter et al., 2020, Rossiter et al., 2021), thesis supervision is a relatively less explored topic. The proposed approach in this paper aims to resolve some of the aforementioned challenges in thesis supervision to establish information accessibility, communication, and collaboration (Karunaratne, 2018). Control engineering has had a significant influence on systems engineering approaches, so perhaps it is possible to benefit from counterpart's prevalent solutions. Hence, this paper proposes a blended systems engineering approach with recent software tools in engineering thesis supervision. Even though the proposed content aims at the control engineering education, a similar context can be adopted in other fields of study.

Further descriptions are provided in upcoming sections. Section 2 addresses the research motivation and the gaps between academic curriculum and industry demand in the field of control engineering. In Section 3, a V-model for control engineering theses progress management is introduced. Section 4 outlines the Agile Scrum approach and subsequently, the proposed Scrum model for thesis supervision is presented in Section 5. The blended process to utilise Agile Scrum, V-model and software tools in postgraduate theses is explained in Section 6. Finally, Section 7 concludes the paper.

## 2. MOTIVATION

Marketplace expectations from a control specialist deviates greatly from working only with poles and zeros or nonlinear models, i.e., pure control theory (Hearns, 2020). Control engineering is a multidisciplinary field that demands hands-on knowledge and experience in a diverse range of topics such as modelling, simulation, implementation, test, integration, software engineering, other engineering subjects, and a profound understanding of the system. Hence, in practice, a control specialist with systems engineering experience is desirable, and relevant learning materials should be envisaged in control education. Systems engineering is a broader integrative view which facilitates project realisation from concepts and requirements definition to implementation, test, verification & validation (V&V), i.e., whole life cycle of a system. Systems engineering is widely influenced by control engineers and related concepts such as closed-loop feedback, optimisation, system view, etc. (Kossiakoff, et al., 2011, Buede and Miller, 2016). Therefore, more effort in improving students' knowledge in systems engineering results in better future opportunities for them.

Figure 1 illustrates a range of disciplines that make up systems engineering knowledge, modified from (Kossiakof, 2011). The topics above the top-dashed line (green) are typically covered in the academic curriculum, whereas industries demand additional skillset and expertise depicted between the two dashed lines. Hence, there are skill gaps between a graduated engineer and the market demand (highlighted in Figure 1). Even though some of this knowledge can be gained in the industry, filling the gap in academic education by embedding systems engineering topics in curriculum and pedagogy design, can empower the students to meet job requirements leading in more job opportunities for them.

On the other hand, exploiting well-established systems engineering approaches could assist the realisation of a thesis by increasing the quality and decreasing the time. Therefore, the application of formal systems engineering approaches is recommended to:

1- Improve the student's practical knowledge and cover the mentioned gap during their study to be industry-ready.

2- Assist them in delivering high-quality thesis projects in a timely manner.

The present study sets out to pave the way for the above-mentioned goals.

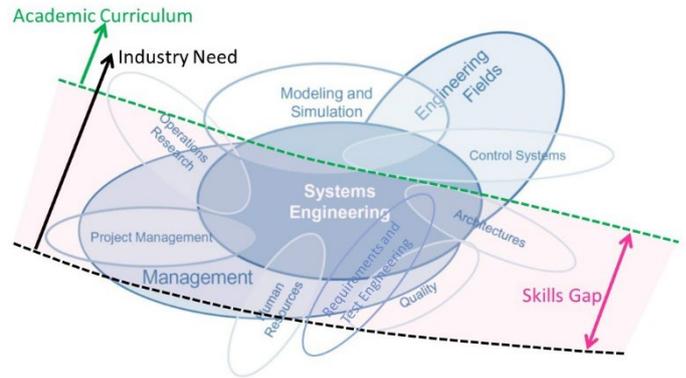

Figure 1. Different expertise related to systems engineering; academic curriculum covers above the top (green) line, industry needs over the bottom (black) line and the skills gap.

## 3. V-MODEL APPLICATION IN CONTROL ENGINEERING THESIS

The V-model has gained substantial interest in diverse engineering topics, more specifically in the software and systems engineering (Weilkiens, et al., 2015, Sarhadi and Yousefpour, 2015). This model is rather a conceptual diagram to illustrate the system development process. Although there are critiques to this model, it is still widely accepted in the systems engineering community (INCOSE, 2015). The V-model's emerging history goes back to the 1960s (Weilkiens, et al., 2015), with similar roots in pedagogy dating back to the 1980s (Gowin and Alvarez, 2005). However, its major popularity has been since 1990 in the software engineering discipline to incorporate the verification and validation processes (Weilkiens, et al., 2015, INCOSE, 2015).

The V-model derives from the core Systems Engineering philosophy of "Define the requirement, then prove your solution meets that requirement". The different levels of the V reflect the recognition that big problems will require complex solutions: systems made up of a series of sub-systems, each one of which may contain multiple hardware and/or software components. The V-model emphasises that the requirements at each level must ultimately be derived from the top-level stakeholder requirements, and each sub-system or component is to be verified against these requirements prior to integration into the whole. Note that different system architectures may lead to different sub-systems and/or components; hence diverse requirements at the lower levels of the V could still meet the same top-level need. Owing to the affinity between control, software, and systems engineering, and the need for effective progress control approaches in thesis supervision, a V-model is proposed. This customised model for control engineering theses is shown in Figure 2.

In Figure 2, the top left-side is dedicated to the conceptual philosophy behind the thesis. Problem(s) to be solved, targets, requirements, and the roadmap of the thesis as well as the milestones should be defined in the first steps. As with any conventional V-model, the top-level determines the main concepts based on the stakeholders' needs. The stakeholders for a thesis are mainly the funding body of the project, the research community, and the academic institution (supervisors and panel). On the other hand, students are inverse stakeholders for the supervisor and the university. Lower levels of the V-model embrace the controller design and implementation procedure. This bottom level mainly includes the MSDIT procedure standing for the first words of the following: 1) **M**odelling; 2) **S**imulation; 3) **D**esign; 4) **I**mplementation; 5) **T**esting, where (1-3) are concepts from classic control theory. However, the implementation and test topics should be emphasised and regularly updated based on the industry demand. On the right-hand side and the top of the V-model in Figure 2, employing the high-fidelity simulators and laboratory tests such as Model/Software/Hardware in the Loop (MIL/SIL/HIL) simulations are recommended. Practical implementation is a desire rather than a goal. In the last stages on the right-hand side, the community feedback over the publications, satisfaction of the requirements and the goals of the thesis, and the knowledge attained by students should be considered. Verification and Validation (V&V) can be performed by matching the left and the right-hand sides of the V-model. Responsibility for V&V is by the supervisor(s), student(s) and other stakeholders. From Figure 2, feedback and review is evident for each block, so the process is iterative, and each block may be amended based on project progress. Thus, the proposed scheme is rather an iterative V-model compared to the classical versions (Kossiakof, 2011).

## 4. AGILE SCRUM DEVELOPMENT: AN OVERVIEW

Agile Scrum (Takeuchi and Nonaka 1986) arose from analysis of the reasons for failure in major software development programmes and evidence that these were often driven by a failure to fully capture or understand the fundamental requirements. This in turn often led to an early commitment to an inappropriate system architecture or design, with erroneous or incomplete requirements flow-down to sub-systems and components. In Systems Engineering terms, Agile Scrum is a rejection of the idea that the V-model is a "one way" or a "one shot" process. Agile Scrum instead proposes an iterative development cycle, with the product built up incrementally across a series of "Sprints". The process can be summarised as follows:

- Capture the key customer requirements, ideally as a set of "User Stories" or "Use Cases", plus any known design constraints. These form the Product (or Project) Backlog.
- Select a sub-set of (high priority) stories to be developed first; these will form the Sprint Backlog. At a Sprint Planning Meeting, the development team plans how to deliver these requirements, with the resulting tasks recorded and added to the Sprint Backlog. The team commits to deliver in a short timescale (typically 1-4 weeks).
- During the sprint, the team meets on a regular basis (the Daily Stand-up or Scrum Meeting) to ensure that they are meeting their commitments and to resolve any issues that may arise.
- At the end of the Sprint, two events are held:
  o Sprint Review: the team demonstrates the delivered product element for acceptance by the product owner. Any refinements to the team's understanding of the requirements are captured in the Product Backlog.
  o Sprint Retrospective: the team reflect on the approach taken to development and propose actions to improve their effectiveness in the future.
- The Sprint cycle is repeated until all accepted requirements in the Product Backlog have been delivered.

The Agile Scrum approach recognises that the team has imperfect knowledge and experience in the early stages of the project and hence emphasises continuous improvement through two feedback mechanisms: requirements are refined at the Sprint Review and captured in the Product Backlog; and the team itself is encouraged to identify better ways of working through time explicitly set aside for the retrospective. Changes to the system design and/or architecture as the project progresses are perfectly acceptable (and even expected) if justified by either customer feedback or team experience.

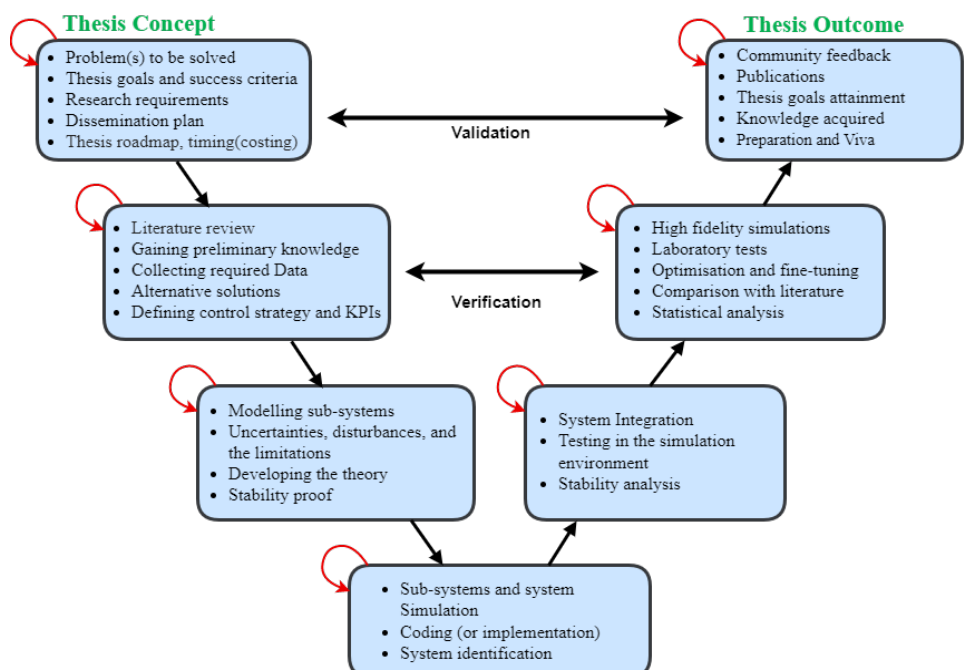

Figure 2. V-model for control engineering thesis progress management

Although the V-model is sometimes (erroneously) seen as being in opposition to Agile Scrum, it should be apparent from the above that they are complementary methodologies. Each User Story delivered in Agile Scrum implies a full run through the V-model, from requirement definition through to user acceptance, with multiple iterations of the Design-Implement-Verify loop potentially occurring throughout the Sprint (see figure see Figure 3).

## 5. AGILE SCRUM PROJECT MANAGEMENT IN CONTROL ENGINEERING THESIS SUPERVISION

Based on the previous section's preface, an Agile Scrum framework for thesis supervision is proposed. There are differences to the basic Agile Scrum and the concept of the team is replaced with supervisor(s) student(s) group (Alblas, 2018). The proposed framework is depicted in Figure 3.

In the thesis definition, a big picture schedule of the project is extracted. The process is to break down the project into smaller short-term tasks (or "tickets") to be conducted in two-week periods, called "Sprints" (Linden, 2018). Regular Sprint meetings are arranged biweekly for half to one hour for each specific student. In the Sprint period, informal email or chat contacts are carried out (e.g., sharing new references in the field, providing feedback, etc). The principle is to define tickets (i.e., the issues to be fixed or tasks to be done in Scrum terminology, Igaki, et. al, 2014) to be investigated in each Sprint. To facilitate the process, at the end of each Sprint and one day before the meeting, students need to prepare a short (one page) report addressing the following questions:

- What has been done since the last meeting?
- What was the key outcome/finding(s)?
- How many hours did you spend in this period?
- What will we do in the next Sprint?
- What are the problems? and what kind of help/support can the supervisor(s) provide?

A technical report regarding the study or simulation may be appended to the progress report. By studying the report(s), supervisor(s) assess the progression of the students and prepare for the meeting. This report provides a self-assessment and feedback framework for student-supervisor. Thus, both of the parties attend to the meeting with a precise background to effectively use the meeting time. During the Sprint meeting, issues are discussed whilst supervisors help in different areas such as: expressing their thoughts, providing further references, support in coding, writing the papers, etc. Based on Figure 3, the process is divided in six main steps (S1-S6) as follows:

S1) Define the task(s):
In this step, the task(s) to be accomplished based on the breakdown of the project roadmap in the next Sprint period are defined. Those tasks can be reviewed during each Sprint.

S2) Create a backlog list:
Once any task or issue is recognised, it should be recorded in a backlog list with adequate descriptions. So, those tasks can be completed in a particular Sprint.

S3) Sprint planning:
Planning for the next Sprint is conducted in the Sprint meetings. Some tasks from the backlog can be nominated to be accomplished in the Sprint period. In this paper, formal biweekly meetings are recommended which could initially be set weekly adapting with students' progress.

S4) Sprint review:
Generally, both S3 and S4 are performed during a Sprint meeting. In each meeting, first, the task(s) considered for the previous Sprint are reviewed. The existing problems are identified and discussed. Then, a plan for the next Sprint and the tasks to be performed are drawn out (Sprint retrospective). Those tasks are named "tickets". The backlog can also be updated based on current findings. By implementing this procedure, at the end of each Sprint increments (steps towards the goals) could be achieved until the end of the thesis.

S5) Informal catchups:
Instead of daily stand-up meetings in conventional Scrum (asking for daily progression), informal catchups during the Sprint are recommended. The authors believe asking for daily feedback could be considered extreme for university research and might be non-pragmatic because of lecturing schedule. However, informal supervisor-student contacts should be held by email, chat, or calls to observe the progression and provide feedback. In addition, the short report mentioned in S5 covers the topics discussed in the conventional Scrum stand-ups.

S6) Repeat: The described procedure should be constantly and regularly repeated until the end of the research.

Due to the nature of Scrum, the proposed method could be easily exploited for large number of students to address the "mixed model" supervision proposed in McCallin and Nayar (2012).

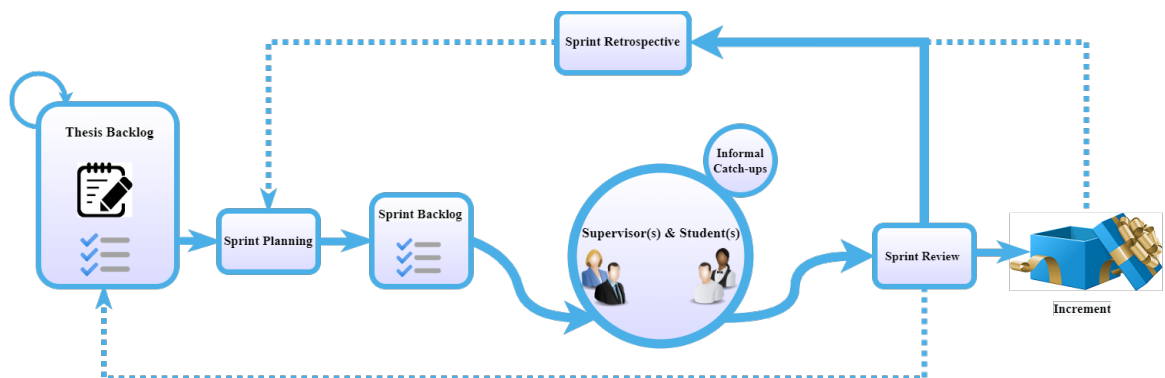

Figure 3. Agile Scrum framework for thesis supervision with supervisor(s)-student(s) affairs

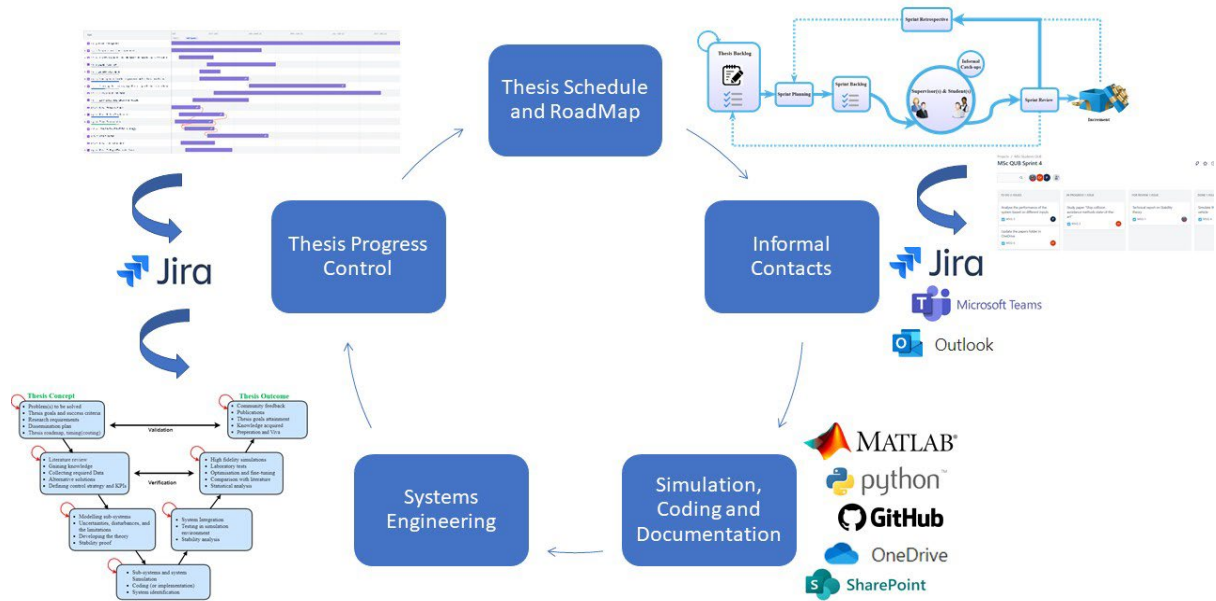

Figure 4. The proposed blended approach to be utilised in thesis supervision

Interestingly, the proposed approach can be interpreted with the general concepts of control systems. For example, regular meetings can be considered as the constant sampling (digital control) of the project. The feedback and feedforward, concepts are realised by the discussions in the Scrum meetings. Backlogs play the role of integrators. Setpoints and outputs are respectively defined on the top (left- and right-hand) blocks of the V-model.

## 6. THE PROPOSED BLENDED APPROACH

Application of Agile approaches in education and pedagogy is proposed in some studies (Igaki, et. al, 2014, Tengberg, 2015, Linden, 2018, Brodtkorb, 2019, Komar, et. al, 2021). However, this paper presents a process to integrate systems engineering aspects, i.e., V-model (Section 3), Agile Scrum process management (Section 4) and state-of-the-art software tools for supervision facilitation. The proposed blended approach is a circular framework illustrated in Figure 4.

Based on Figure 4, the procedure starts from the thesis roadmap and breakdown on the top level. Agile Scrum is utilised to proceed with the research. Jira software from Atlassian, Jira (2022), is proposed to implement the Agile Scrum introduced in Section 5. Although the main intention of Jira is project management, it provides compelling features for educational purposes (Scott and Campo, 2021). This application realises an online Scrum framework that contains roadmaps, calendar, backlogs, tickets, Sprint meetings arrangement, i.e., whatever is required in an Agile Scrum implementation. Jira allows defining multiple projects-teams, and it is free to use for small groups (<10 members). Furthermore, Jira provides graphical boards reflecting the tasks for a particular Sprint, and their status in different: "To Do", "In Progress", "For Review", and "Done" columns to easily monitor the progression. A typical Jira board with the aforementioned columns is shown in Figure 5. Overall, employing this powerful tool could be a big step forward to realise Agile Scrum in thesis supervision.

Besides, informal catchups and contacts can be held by MS Teams calls and chat (Microsoft Teams, 2022) as well as emails that are compatible with online education requirements in CoVid19 pandemic restrictions.

Control engineering is tightly interconnected with programming and simulations tools where MATLAB and Python are the commonly used coding frameworks. Nevertheless, in this paper, exploiting version control techniques such as Git are recommended instead of other code sharing methods. As an instance, GitHub (2022) provides an online environment for collaborative development and version control. Further to this, Git is fast becoming one of the essential skills to secure a job, so learning this tool could make students more competitive on the job market (Hsing and Gennarelli, 2019). For sharing data and reports, Microsoft tools such as OneDrive or SharePoint can be used because of their accessibility in most universities. Otherwise, similar applications such as Slack, Confluence, GitLab, Bitbucket, etc. could be replaced to close the cycle proposed in Figure 4. In addition, there are options to integrate Jira with other tools such as GitHub, MS-Teams, etc. Finally, the V-model introduced in Section 3 is proposed to be used to close the cycle by defining the requirements and establishing the V&V process (Figure 4).

Therefore, application of this blended approach could be a key step towards productive thesis supervision and enhancing the students' skills in the "university to market" gap area.

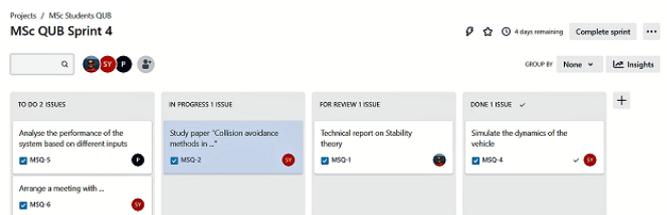

Figure 5. A typical Jira board with tasks' status.

# 7. CONCLUSIONS

The paper presents a blended approach and relevant tools to support the supervision of postgraduate student theses. Firstly, some of the current skill gaps between university education and job market demand for graduate control engineers are discussed. To fill this gap, and facilitate the student's research, a chain of systems engineering techniques and recent software solutions are proposed. A customised V-Model for thesis accomplishment is presented which considers the whole life cycle and relevant stakeholders. An Agile Scrum process to define and follow up sub-tasks supporting regular Sprint meetings with feedback and self-assessment reports is recommended. Finally, a blended approach to implement the aforementioned concepts with available software applications is addressed. The proposed framework is designed to be compatible with online education requirements during the CoVid19 pandemic restrictions. Furthermore, with some modifications, the general idea of the Agile approach could also be applied to purely theoretical projects. Students' progress has been enhanced through the application of the proposed approach; Nevertheless, the proposal is still in its infancy and requires further investigation.